\documentclass[twocolumn,prb,showpacs]{revtex4}
\usepackage{graphicx}

\begin{document}
\title{Biconical structures in two--dimensional anisotropic Heisenberg
antiferromagnets}
\author{M. Holtschneider}
\author{W. Selke}
\affiliation{Institut f\"ur Theoretische Physik B,
 RWTH Aachen, 52056 Aachen, Germany}

\begin{abstract}

Square lattice Heisenberg and XY antiferromagnets with uniaxial anisotropy in
a field along the easy axis are studied. Based on ground state
considerations and
Monte Carlo simulations, the role of biconical structures 
in the transition region between the antiferromagnetic
and spin--flop phases is analyzed. In
particular, adding a single--ion anisotropy to the XXZ 
antiferromagnet, one observes, depending on the sign of that
anisotropy, either an intervening biconical phase  or a
direct transition of first order separating the two phases. In case of
the anisotropic XY model, the degeneracy of the ground state, at
a critical field, in antiferromagnetic, spin--flop, and bidirectional
structures seems to result, as in the case of the XXZ model, in
a narrow disordered phase between the antiferromagnetic and spin--flop
phases, dominated by bidirectional fluctuations.
\end{abstract}


\maketitle

Recently, two--dimensional uniaxially anisotropic Heisenberg
antiferromagnets in a magnetic
field along the easy axis have been studied theoretically rather
intensively \cite{Mat,Leidl1,Holt1,Zhou1,Leidl2,Holt2,Vic,Schwing,Zhou2}, 
motivated by experiments on intriguing magnetic properties of
layered cuprates \cite{Mat,Ammer,Pokro,Rev1,Rev2} and by experimental
findings on complex phase diagrams for other quasi two--dimensional
antiferromagnets \cite{Bevaart,Gaulin,Cowley,Chris,Pini} exhibiting, 
typically, multicritical behavior.

A generic model describing such systems is the XXZ Heisenberg
antiferromagnet on a square lattice, with the Hamiltonian 

\begin{equation}
{\cal H} = J \sum\limits_{(i,j)} 
  \left[ \Delta (S_i^x S_j^x + S_i^y S_j^y) + S_i^z S_j^z \right] - 
  H \sum\limits_{i} S_i^z 
\end{equation}

\noindent
where we consider the classical variant, with
the spin at site $i$, $\vec{S}_i=[S_i^x,S_i^y,S_i^z]$, being a vector 
of length one. $\vec{S}_i$ is coupled to its four neighboring spins 
$\vec{S}_j$ at sites $j$. The exchange integral $J$ is
antiferromagnetic, $J>0$, and the anisotropy
parameter $\Delta$ may vary from zero (Ising limit)
to one (isotropic Heisenberg model). The magnetic field $H$ acts along
the easy axis, the $z$--axis. As known for many years \cite{LanBin},
the phase diagram of the XXZ model includes the long--range ordered 
antiferromagnetic (AF), the algebraically ordered spin--flop (SF), and
the paramagnetic phases. Only very recently, attention has been
drawn to the role of biconical (BC) structures and fluctuations, in
the ground state and in the transition region between the AF and
SF phases \cite{Holt2}.

In a BC ground state configuration the spins on the two 
sublattices (i.e. on neighboring sites), $A$ and $B$, form different
cones around
their two different tilt angles, $\theta_A$ and $\theta_B$, with respect
to the easy axis, see Fig.\ 1. In the XXZ model, these structures occur at
the critical field, $H_{c1}$, which separates the AF and SF
structures at $T=0$. The two tilt angles of the BC ground
states are interrelated by \cite{Holt2}.

\begin{equation}
\theta_B = \arccos \left( \frac{ \sqrt{1-\Delta^2} \; - \; \cos\theta_A }{ 1 \; - \; \sqrt{1-\Delta^2} \cos\theta_A } \right) 
\end{equation}

\noindent
leading, in addition to the rotational symmetry in the xy-components
of the spins in the BC and SF states, to a high degeneracy of the
ground state. This degeneracy
seems to give rise to a narrow intervening, possibly disordered phase in the
(field $H$, temperature $T$)--phase diagram of the square lattice XXZ
model, as discussed before \cite{Holt2,Holt1,Zhou1}.

\begin{figure}
  \begin{center}
    \includegraphics[width=0.95\linewidth]{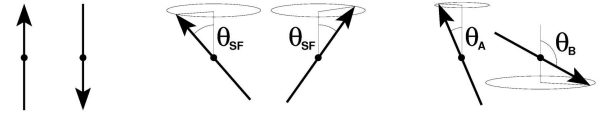}
  \end{center}
  \caption{\label{fig1} Spin orientations on neighboring sites
    in the ground states of the XXZ models (without and with
    additional single--ion anisotropy), corresponding to, from left
    to right, antiferromagnetic, spin--flop, and biconical configurations.}
\end{figure}

In this communication, we study variants of the XXZ model, staying
in two dimensions, by adding a single--ion anisotropy, and by reducing the
XXZ to the anisotropic XY model. Results will be based on ground state
calculations and Monte Carlo simulations. The aim is to analyze the effect 
of these modifications on the biconical (or their 'bidirectional'
analogues for the XY
variant) structures and fluctuations, both
for the ground states and for the phase diagrams.
 
Biconical structures can be
stabilized in various ways, starting with the XXZ model, especially
by taking into account further anisotropies, for instance, a cubic
anisotropy, or couplings to spins at more distant
sites \cite{Tsun,Fisher,Bruce}. Here, a single--ion
anisotropy will be added, reflecting crystal
symmetry. Similar properties are expected for related anisotropies as
well. The single--ion anisotropy term has
the form

\begin{equation}
{\cal H}_{si} = D \sum\limits_{i} (S_i^z)^2
\end{equation}

\noindent
which either, $D < 0$, enhances the uniaxial
anisotropy $\Delta$, or, $D>0$, may weaken it due to a
competing planar anisotropy. The sign of $D$ will have 
drastic consequences for the phase diagram. In both cases, the ground state
properties can be determined exactly \cite{Holt3}.

In case of a {\it positive} single--ion anisotropy, $D>0$, BC structures
are ground states in a non--zero range of
fields, $H_{c1a}< H< H_{c1b}$, in between the
AF ($H \leq H_{c1a}$) and SF ($H \geq H_{c1b}$)
structures \cite{Holt3}. Note that
at fixed field in that range, only BC structures with a unique pair
of tilt angles $\theta_A$ and $\theta_B$, are stable, with a modified
relation between the angles, compared to Eq. (2), depending now
also on $D$.  At $T>0$, there is a BC phase, ordered simultaneously
in the spin components
parallel and perpendicular to the easy axis \cite{Fisher}, intervening
between the
AF and SF phases. This is found in extensive Monte 
Carlo simulations studying square lattices with up to 
$L \times L = 240 \times 240$ sites, and
performing runs with up to $10^8$ Monte Carlo steps per site (MCS), applying
the standard Metropolis algorithm. A typical phase diagram is shown    
in Fig. 2, where we set $\Delta= 0.8$, as
usual \cite{LanBin,Holt1,Zhou1}, and $D/J= 0.2$. The phase boundaries
are determined by finite--size extrapolations of the
staggered susceptibilities, the specific heat, and the Binder
cumulant \cite{Holt3}. Obviously, see
Fig. 2, the extent of the BC phase shrinks
with increasing temperature. Eventually, the BC phase may terminate at
a tetracritical point \cite{Fisher,Bruce,Muka,Hu,Aharony,Cala}, where
the AF, SF, BC and paramagnetic phases meet. It may be estimated to be
roughly at $k_BT/J = 0.35 \pm 0.05$.

\begin{figure}
  \begin{center}
    \includegraphics[width=0.95\linewidth]{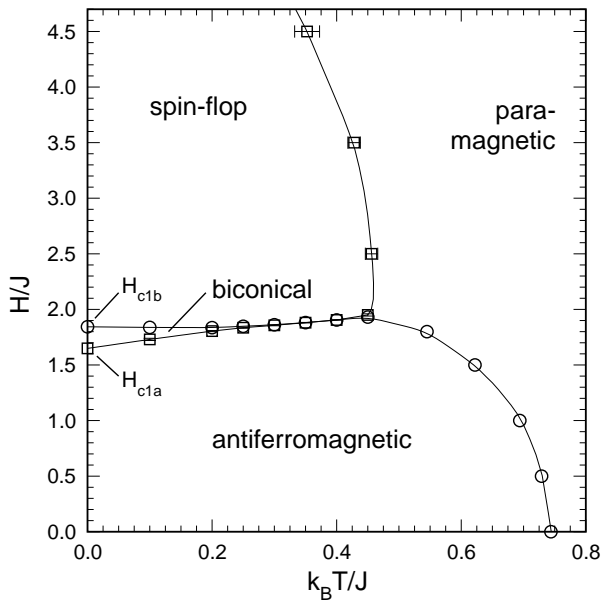}
  \end{center}
  \caption{\label{fig2} Simulated phase diagram of the XXZ
    model, $\Delta= 0.8$,  with an additional
    positive single--ion anisotropy, $D=0.2$. The squares denote the
    SF, the circles the AF phase boundary. Error bars are
    omitted unless they are larger than the symbol sizes.
   }
\end{figure}

The critical properties of the boundary lines of the BC phase to
the AF and SF phases have been simulated and studied in much detail
at $k_BT/J= 0.2$, see
Fig. 2. The boundary line between the BC and SF phases, where the staggered
magnetization in the direction of the field (or 'longitudinal
staggered magnetization') vanishes, seems to belong to
the Ising universality class. For example, the effective critical exponent of
that susceptibility is found to approach 7/4, when analyzing the
size dependence of the peak height \cite{Holt3}. Note that, in this
case, large systems, $L \geq 120$, are needed for getting close
to the supposed asymptotics. In turn, at 
the boundary line between the BC and AF phases the algebraic order
in the transverse staggered magnetization, which we observe in the 
BC phase, gets lost. The finite-size behavior of that
magnetization, for $L \geq 40$,
agrees with the transition belonging to the
Kosterlitz--Thouless universality class. Note that renormalization group
arguments on the universality classes of the boundary lines of
biconical phases \cite{Bruce} suggest transitions in the Ising
and XY--universality classes as well. 
 
In case of a {\it negative} single-ion anisotropy, $D<0$, at all
fields, no BC structures
are ground states. At low temperatures, the
Monte Carlo simulations (with computational efforts as for 
$D>0$) provide evidence for a direct transition of first order
between the AF and SF phases. Such
evidence is exemplified in Fig. 3 for $\Delta = 0.8$ and $D/J= -0.2$,
where the peak height of the longitudinal staggered
susceptibility, $\chi_{max}$, is shown to grow
with size $L$ proportionally to $L^2$, as expected for 
a transition of first order. The coexistence of the AF and SF phases
at a first--order transition is also seen in the behavior of the
probability $P(\theta)$ to find the tilt angle
$\theta$ in a configuration. $P(\theta)$ shows more and more pronounced 
maxima at the values of $\theta$ characterizing the AF and SF
phases, when increasing the system size. Note that for small system
sizes, biconical fluctuations are observed in the transition region between
the AF and SF phases \cite{Holt3}.
           
\begin{figure}
  \begin{center}
    \includegraphics[width=0.95\linewidth]{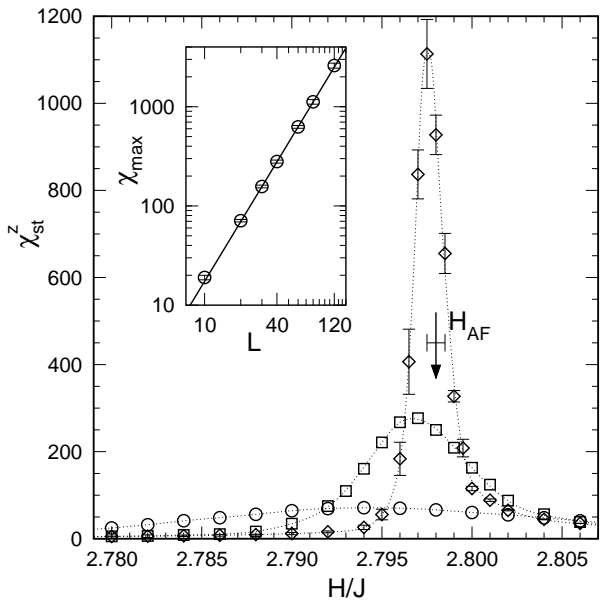}
  \end{center}
  \caption{\label{fig3} Selected raw data for the longitudinal staggered
    susceptibility $\chi _{st}^z$ versus field at fixed
    temperature, $k_BT/J=0.3$ near the
    transition, $H_{AF}/J= 2.798 \pm 0.0005$, between
    the AF and SF phases for the XXZ
    model, $\Delta=0.8$, $D=-0.2$. Data for Systems of sizes 20 (circles), 40
    (squares), and 80 (diamonds) are shown. In the inset, the
    finite--size, $L$, dependence of the peak height, $\chi_{max}$, is
    depicted, with the straight, solid line 
    showing $\chi_{max} \propto L^2$. Error bars are included when they
    are larger than the symbol sizes. 
}
\end{figure}

Let us turn to another variant of the XXZ model, the anisotropic
XY antiferromagnet. According to renormalization group
calculations \cite{Koster,Cala,Vic}, applied to
the case of uniaxiality, the 
number of spin components $n$ is expected to determine
the nature of the multicritical
point, at which the AF, SF, and, possibly, BC phases meet with the
paramagnetic phase. For $n$ being not too large, the multicritical point has
been supposed to be a bicritical, tetracritical or
critical end--point. Now, in two dimensions, a bicritical point
of O(n)--symmetry with $n=3$, as it is the case in the XXZ model, is
ruled out at a non-zero temperature in two dimensions by
the well--known Mermin--Wagner theorem. However, it may
occur when reducing the number of spin components from $n=3$ to 
$n=2$. Then the  bicritical
point would be of O(2)--symmetry, and thus would be allowed
at $T>0$, belonging to the 
Kosterlitz--Thouless universality class. Thence, it looks interesting to
consider the $n=2$ variant of the XXZ antiferromagnet, namely
the anisotropic XY model, described by the Hamiltonian

\begin{equation}
{\cal H} = J \sum\limits_{(i,j)} 
  \left[ (S_i^x S_j^x + \Delta S_i^y S_j^y) \right] - 
  H \sum\limits_{i} S_i^x 
\end{equation}

\noindent
with the classical spins having now only two components. 

The ground state analysis can be done in complete analogy to the one
for the XXZ model. Of course, the spins are now restricted
to the XY--plane, and the tilt angle $\theta$ is defined for the 
spin orientation with respect to the easy $x$--axis. Especially, the
biconical structures are replaced by bidirectional (BD)
structures. At the critical field, $H_{c1}$, the AF and SF structures
are degenerate with the set of BD configurations, for which $\theta_A$ and
$\theta_B$ are interrelated as in the XXZ case, Eq. (2).    

\begin{figure}
  \begin{center}
    \includegraphics[width=0.95\linewidth]{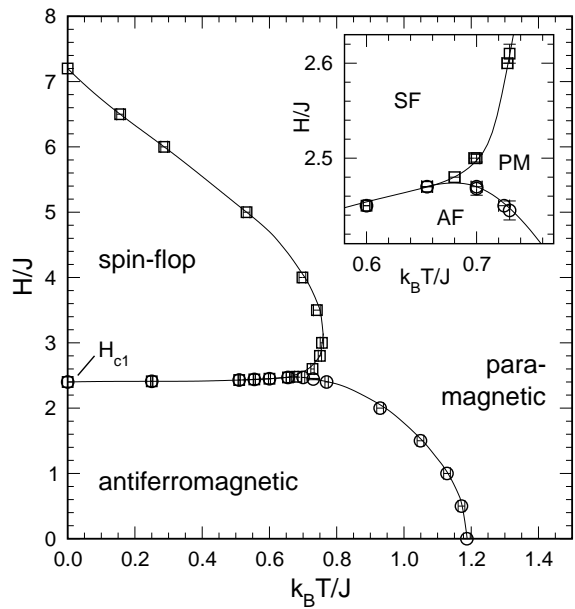}
  \end{center}
  \caption{\label{fig4} Simulated phase diagram of the anisotropic
    XY model, with $\Delta= 0.8$. The squares denote the phase
    boundary of the SF, the circles the one of the AF phase. The inset
    magnifies the part, where AF and SF phase boundaries approach 
    each other.}
\end{figure}

The phase diagram of the anisotropic XY antiferromagnet, with
$\Delta$= 0.8, is depicted in Fig. 4. It has been obtained from
extensive Monte Carlo simulations studying lattices
sizes up to $L \times L=120 \times 120$, and performing runs with up to $10^8$
MCS. The phase boundaries are determined by finite--size
extrapolations, similar to the analysis for the XXZ model with
a single--ion anisotropy.

The topology of the phase diagram looks like in the XXZ 
case \cite{LanBin,Holt1,Zhou1}. The AF and SF boundary lines approach
each other very closely near the maximum of the SF phase boundary in the
$(T,H)$--plane, see Fig. 4. Accordingly, at low temperatures, there
seems to be either a direct transition between the AF and
SF phases, or two separate transitions with an extremely
narrow intervening phase may occur.

Away from that intriguing transition region, one expects the
transition not only from the AF but also the one from the SF phase
to the paramagnetic phase to be
in the Ising universality class. In the SF phase of the XY
antiferromagnet, there is just one ordering
component, the $y$--component. The expectation
is confirmed by the Monte Carlo data for the specific
heat (where the peak at the
AF phase boundary gets rather weak on approach to the transition
region) and for the staggered 
susceptibilities. The quantities exhibit critical behavior of Ising--type, as
follows from the corresponding effective exponents describing size 
dependences of the various peak heights.

\begin{figure}
  \begin{center}
    \includegraphics[width=0.9\linewidth]{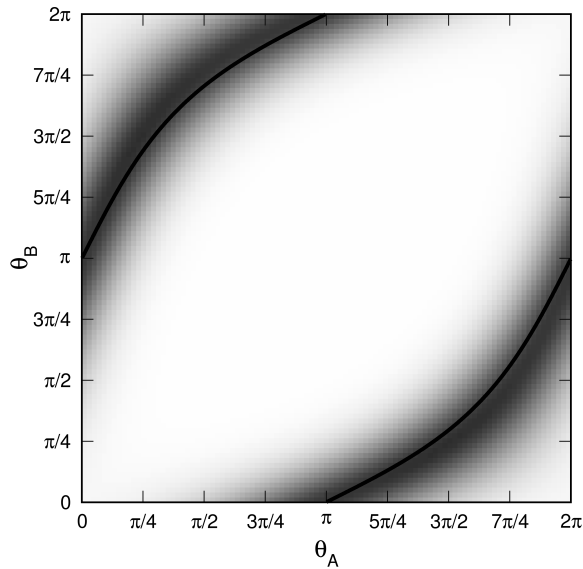}
  \end{center}
  \caption{\label{fig5} Joint probability $p(\theta_A,\theta_B)$ for
    the anisotropic XY antiferromagnet with $\Delta$= 0.8 for a system
    with $100 \times 100$ lattice sites in the transition region between
    the AF and SF phases at $k_BT/J= 0.558$ and
    $H/J=2.44$. $p(\theta_A,\theta_B)$ is proportional to the
    grayscale. The superimposed solid line depicts the relation between
    the two tilt angles in the ground state, see Eq. (2).
}
\end{figure}

In the transition region of the AF and SF phases, BD fluctuations dominate, as
one may conveniently infer from the joint probability distribution
$p(\theta_A,\theta_B)$ for finding the tilt angles $\theta_A$ and
$\theta_B$ at neighboring sites, i.e. for the two different
sublattices. A typical
result is depicted in Fig. 5, showing the behavior of $p$ in a
grayscale representation. Two
features are of interest: first, the two tilt angles are strongly
correlated like in the degenerate ground state, with a line of local
maxima in $p$ closely following Eq. (2). Second, all those bidirectional
structures occur simultaneously with (almost) equal probability, i.e. along 
the line of maxima $p$ is (almost) constant. Note that this behavior
is only weakly affected by finite--size effects for the
system sizes we simulated. Moving away from the
transition region, for instance, by fixing the
temperature and varying the field, the line of local maxima initially
does not change significantly, but, along that line, pronounced 
peaks start to show up at positions corresponding to the
AF, at lower fields, or
to the SF phases, at higher fields. Similar observations for the joint
probability distribution hold for the XXZ model without
single--ion anisotropy \cite{Holt2}.

Additional evidence on critical phenomena in the transition region
may be obtained by analyzing effective exponents. We did that for
the staggered susceptibilities and the specific heat. Results
(especially, at $H/J=2.44$ and $0.54 < k_BT/J < 0.57$) are
compatible with Ising--type criticality, but rather large corrections
to scaling had to be presumed. For example, the effective critical
exponents for describing the size--dependences of the peak height for
the staggered susceptibilities are about
1.8 to 1.85, largely independent of system size. The supposedly
strong corrections to scaling may be due to very large correlation 
lengths in that region, and the asymptotics  may be reached
only for very large systems.

To conclude, the simulational data seem to
suggest for the anisotropic XY antiferromagnet the existence of an extremely
narrow, presumably, disordered phase, intervening between the AF
and SF phases, like in the XXZ case \cite{Holt1,Zhou1}. The transition
region between the two phases is clearly dominated by, in the
ground state completely degenerate, bidirectional fluctuations.

\acknowledgments
We thank Amnon Aharony for useful correspondence. Financial support
by the Deutsche Forschungsgemeinschaft under grant 
~SE~324/4 is gratefully acknowledged.

\end{document}